\documentclass[journal,twoside,web]{ieeecolor}
\usepackage{generic}
\usepackage{multirow}
\usepackage{soul}
\usepackage{float}
\usepackage{cite}
\usepackage{graphicx}
\graphicspath{{./figures/}}
\usepackage{subcaption}
\usepackage{braket}
\usepackage{amsmath,amssymb,amsfonts}
\usepackage{algorithmic}
\usepackage{graphicx}
\usepackage{algorithm,algorithmic}
\usepackage{hyperref}
\hypersetup{hidelinks=true}
\usepackage{textcomp}
\usepackage{xcolor}
\def\BibTeX{{\rm B\kern-.05em{\sc i\kern-.025em b}\kern-.08em
    T\kern-.1667em\lower.7ex\hbox{E}\kern-.125emX}}
\begin{document}
\title{Quantum Machine Learning for Energy-Efficient 5G-Enabled IoMT Healthcare Systems: Enhancing Data Security and Processing}
\author{Muhammad Zeeshan Riaz, Bikash~K.~Behera, Shahid Mumtaz, Saif Al-Kuwari, and Ahmed~Farouk
\thanks{Muhammad Zeeshan Riaz is with the College of Physics and Optoelectronic Engineering, Shenzhen University, Shenzhen 518060, China, e-mail: (zeeshanriaz@email.szu.edu.cn ).}
\thanks{B.~K. Behera is with the Bikash's Quantum (OPC) Pvt. Ltd., Mohanpur, WB, 741246 India, e-mail: (bikas.riki@gmail.com).}
\thanks{Saif Al-Kuwari is with the Qatar Center for Quantum Computing, College of Science and Engineering, Hamad Bin Khalifa University, Doha, Qatar. e-mail: (smalkuwari@hbku.edu.qa).}
\thanks{Shahid Mumtaz is with Nottingham Trent University, Engineering Department, United Kingdom. e-mail: (dr.shahid.mumtaz@ieee.org).}
\thanks{A.~Farouk is with the Qatar Center for Quantum Computing, College of Science and Engineering, Hamad Bin Khalifa University, Doha, Qatar and with the Department of Computer Science, Faculty of Computers and Artificial Intelligence, Hurghada University, Hurghada, Egypt. e-mail: (ahmedfarouk@ieee.org).}}
\maketitle
\begin{abstract}
Energy-efficient healthcare systems are becoming increasingly critical for Industry 5.0 as the Internet of Medical Things (IoMT) expands, particularly with the integration of 5G technology. 5G-enabled IoMT systems allow real-time data collection, high-speed communication, and enhanced connectivity between medical devices and healthcare providers. However, these systems face energy consumption and data security challenges, especially with the growing number of connected devices operating in Industry 5.0 environments with limited power resources. Quantum computing integrated with machine learning (ML) algorithms, forming quantum machine learning (QML), offers exponential improvements in computational speed and efficiency through principles such as superposition and entanglement. In this paper, we propose and evaluate three QML algorithms, which are $UU^{\dagger}$, variational $UU^{\dagger}$, and $UU^\dagger$-\ quantum neural networks (QNN) for classifying data from four different datasets: 5G-South Asia, Lumos5G 1.0, WUSTL EHMS 2020, and PS-IoT. Our comparative analysis, using various evaluation metrics, reveals that the $UU^\dagger$-QNN method not only outperforms the other algorithms in the 5G-South Asia and WUSTL EHMS 2020 datasets, achieving 100\% accuracy, but also aligns with the human-centric goals of Industry 5.0 by allowing more efficient and secure healthcare data processing. Furthermore, the robustness of the proposed quantum algorithms is verified against several noisy channels by analyzing accuracy variations in response to each noise model parameter, which contributes to the resilience aspect of Industry 5.0. These results offer promising quantum solutions for 5G-enabled IoMT healthcare systems by optimizing data classification and reducing power consumption while maintaining high levels of security even in noisy environments.
\end{abstract}
\section{Introduction}
\IEEEPARstart{T}{he} Internet of Medical Things (IoMT) is a crucial application of Industry 5.0 in healthcare, connecting medical devices and patients to revolutionize healthcare delivery globally \cite{guo2021hybrid}. By establishing a robust infrastructure between medical software and hardware applications, IoMT facilitates seamless interaction between biosensor nodes \cite{ahmed2024insights} and mobile edge computing (MEC) \cite{awad2022utilization}, linking patients with doctors and sharing data through secure networks, thus minimizing hospital visits and reducing workload on healthcare departments. The development of 5G networks accelerates the evolution of the Internet of Things (IoT) and IoMT within the Industry 5.0 healthcare system, supporting technologies such as device-to-device (D2D) communication \cite{suraci2021mec}, machine-to-machine (M2M) interactions \cite{ghubaish2020recent}, and mobile cloud communications (MCC) \cite{bolivar2018deployment}. 5G networks ensure fast and secure data transfer between healthcare systems and patients, offering significant advantages over low-bandwidth networks. IoMT services require minimal data usage, energy efficiency, and secure connections \cite{ahad20246g}. 5G technology meets these needs by improving data management and connectivity and providing advanced capabilities such as imaging and treatments. This ensures that multiple doctors can access patient data for diagnosis and decision making, improving service quality. Despite these advancements, establishing a reliable IoMT with 5G for Industry 5.0 infrastructure presents challenges. The safety hazards of wireless healthcare equipment, such as ensuring energy-efficient transmission for wireless body area networks (WBANs), are discussed in \cite{sodhro2016energy}. Energy transmission efficiency and security concerns have emerged as significant difficulties. The expected adoption of IoMT is delayed by users' lack of security awareness and potential vulnerabilities that could expose sensitive health-related information. Overcoming these energy and security challenges is critical to ensure the successful and secure use of IoMT in the healthcare industry 5.0 \cite{pradyumna2024empowering}.

Integrating machine learning (ML) to develop IoMT with 5G for Industry 5.0 healthcare management systems provides significant benefits, including enhanced healthcare quality control, accuracy, effective monitoring, and improved treatment during emergencies \cite{dong2020edge}. ML algorithms facilitate large data transfers, allow devices to share data in real time with telemedicine applications, and allow remote patient monitoring. Various ML techniques improve the efficiency of Industry 5.0 healthcare applications by handling large amounts of data in the initial stages, such as biomedical imaging analysis. Integration of federated learning (FL) with artificial intelligence (AI) in intelligent healthcare systems addresses security, privacy, stability, and reliability issues, offering novel FL-based AI applications in intelligent healthcare, including electronic health record (EHR) management, health monitoring, biomedical image analysis, and COVID-19 trait identification \cite{nguyen2022federated}.  Implementing deep reinforcement learning for secure data transfer minimizes energy consumption, improving battery life compared to greedy algorithms; however, transferring excess data without the patient's consent can quickly drain the battery \cite{allahham2020see}. An innovative approach combining Bell's inequality with chaotic maps has produced 256-bit keys for biosignals and medical imaging through symmetric encryption strategies, and the efficiency is tested for breast tumor detection and biosignal analysis through convolutional neural networks (CNN) \cite{chen2024information}. Furthermore, unsupervised ML methods employ clustering and dimensionality reduction approaches, which are critical to identify hidden patterns, improve patient profiles, and group documentation \cite{baker2023artificial}. Several deep learning and ML techniques enhance patient data safety and monitor healthcare systems in the autism center for 5G networks linked to IoMT \cite{q_17}.

Quantum Machine Learning (QML) has the potential to revolutionize precision in Industry 5.0 healthcare by offering real-time solutions through pattern recognition, task execution, modification of healthcare features, and enhancement of diagnostic and treatment plans, which outperform the effectiveness of classical ML approaches. QML algorithms have been used to improve disease prediction and classification \cite{q_3}, and a hybrid classical-quantum algorithm using clustering of quantum k-means (qk-means) has been proposed to analyze breast cancer and knee magnetic resonance datasets \cite{q_15}. Another study ensures data confidentiality using quantum-enhanced data preservation techniques for smart healthcare systems \cite{laxminarayana2022quantum}. To address the challenges of conventional ML, such as long training times, infrequent output, and security issues, a digital twin-assisted quantum federated learning (DTQFL) approach has been proposed. Digital twins are created using DTQFL over 5G networks for specific patient diseases and are synchronized with quantum variation neural networks to train and update without compromising real-world performance \cite{qu2023dtqfl}.
\subsection{Problem Statement}
Although classical ML methods have improved the security and energy efficiency of 5G-enabled IoMT networks for Industry 5.0, they face significant limitations. These include difficulties handling massive data transfers, ensuring secure data transmission, addressing scalability challenges, and operating efficiently in heterogeneous high-speed environments. QML offers a promising solution to process complex high-dimensional data more efficiently than classical methods, improving scalability, security, and efficiency in large-scale 5G-enabled IoMT deployments. However, despite these advantages, QML algorithms remain highly sensitive to quantum noise, which can introduce security vulnerabilities in patient data and significantly degrade system performance if not adequately mitigated.
\subsection{Novelty}
To fill these critical gaps, we propose a novel QML technique, the $UU^\dagger$-QNN. This method combines the strengths of the $UU^{\dagger}$ algorithm and Quantum Neural Networks (QNN) to improve energy efficiency, scalability, and security in 5G-enabled IoMT for Industry 5.0, while demonstrating robustness against various noisy channels. Unlike conventional QML approaches, which rely on parameterized quantum circuits and suffer from barren plateau problems and vanishing gradient-cost functions when scaling to large numbers of qubits, the proposed $UU^\dagger$-QNN leverages unitary operations and adjoint transformations to stabilize training and preserve quantum information. The proposed algorithm constructs its cost function based on the inner product of $U$ and $U^\dagger$, maintaining unitarity throughout the optimization process. This design reduces the complexity of landscape optimization and enables a more stable and efficient training of quantum models compared to traditional QNNs. Furthermore, $UU^\dagger$-QNN addresses noise-related challenges, making it practical for real-world deployment in 5G-enabled IoMT environments. Experimental evaluation in four benchmark datasets: 5G-South Asia (5G-SA), Lumos5G 1.0 (L5G1.0), WUSTL EHMS 2020 (WE20), and Privacy and Security
Internet of Things (PS-IoT) versus $UU^\dagger$, variational-$UU^\dagger$, and classical models show that $UU^\dagger$-QNN outperforms existing methods, particularly in handling large-scale medical data under noisy conditions.

\subsection{Contributions}
\begin{itemize}
\item[1)] We introduce a novel algorithm $UU^{\dagger}-QNN$ that combines the $UU^{\dagger}$ technique with QNN within a variational quantum circuit framework. This approach improves data security, processing efficiency, and energy optimization in 5G-enabled IoMT systems for Industry 5.0.
\item[2)] We conduct a comprehensive comparative analysis of three quantum classification algorithms $UU^{\dagger}$, variational $UU^{\dagger}$, and the proposed $UU^{\dagger}$-QNN on four diverse datasets: 5G-SA, L5G1.0, WE20, and PS-IoT.
\item[3)] We rigorously evaluate the robustness of the proposed algorithms against five noise models, demonstrating their resilience and effectiveness in noisy environments.
\end{itemize} 

\subsection{Organization}
The paper is structured as follows: Section \ref{SecII} outlines the problem formulation and methodology for quantum algorithms ($UU^\dagger$, variational $UU^\dagger$, and $UU^\dagger$-QNN). Section \ref{SecIII} presents the experimental results for the four datasets considering noise and noiseless environments. Section \ref{SecIV} discusses the experimental results and concludes the study.
\begin{figure*}[]
\centering
\includegraphics[width=0.9\linewidth]{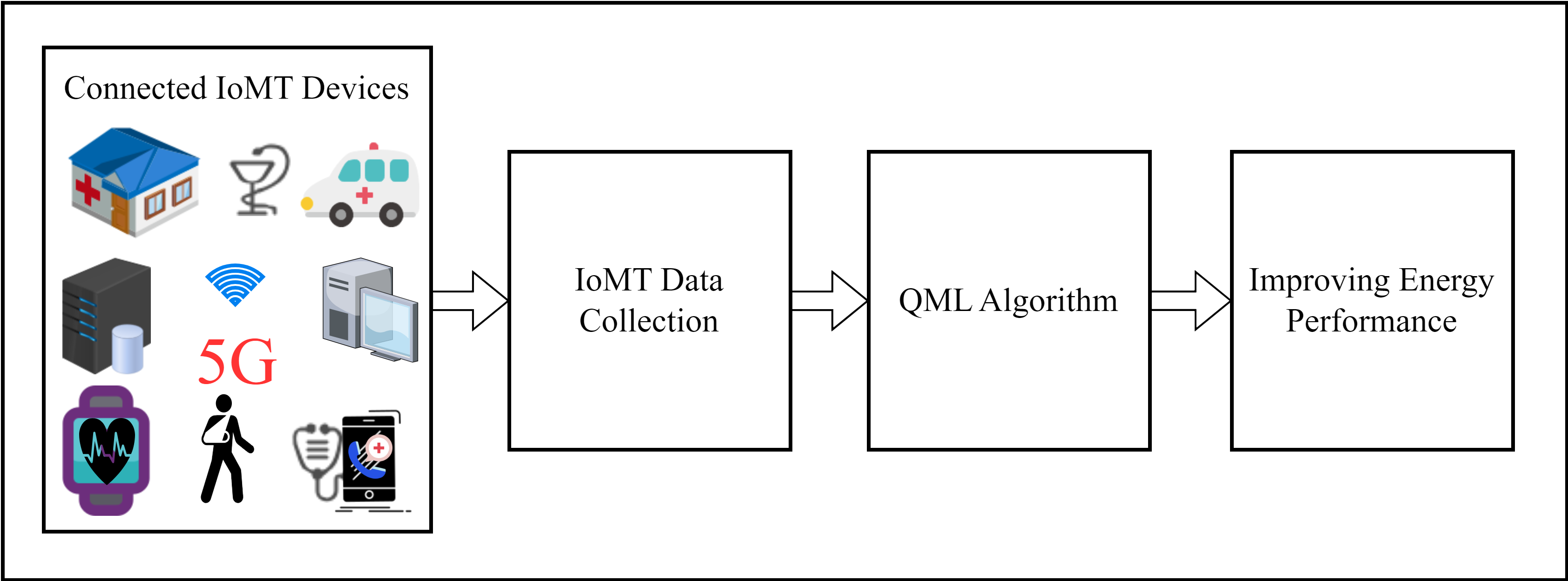}
\caption{Quantum Machine Learning for 5G-enabled IoMT System.}
\label{IoMT_QML.png}
\end{figure*}
\section{Methodology \label{SecII}}
The application of QML techniques for data classification in IoMT devices connected through 5G networks is shown in Fig. \ref{IoMT_QML.png}. IoMT chips integrated into various medical instruments perform early-stage data analysis. This early analysis enables efficient classification tasks using QML techniques, significantly improving the energy performance of connected IoMT devices.

\subsection{Problem Formulation}
Before applying the proposed classical-quantum classification approach, datasets undergo preprocessing steps, including data cleaning, feature scaling, categorical data encoding, and feature selection. The original dataset is represented as:
\begin{eqnarray}
X=(X_1, X_2........,X_M)
\end{eqnarray}
which, after preprocessing, becomes:
\begin{eqnarray}
\overline{X}=(\overline{X_1}, \overline{X_2},........,\overline{X_N})   
\end{eqnarray}
where $\overline{X_i}$ denotes a scaled feature value, and $N\leq M$. The preprocessed dataset consists of $\mathbb{P}$ data points (rows) and $N$ features (columns). In quantum classification algorithms, specific quantum circuits are executed. The condition determines the number of qubits $m$ required for these quantum algorithms:
\begin{eqnarray}
2^m \geq N
\label{qubits}
\end{eqnarray}
After preprocessing, $k$-means clustering is applied to divide the dataset into $k$ clusters: $C_0, C_2,....,C_{k-1}$. The datasets given are divided into two groups, namely $C_0$ and $C_1$. The centroids of these clusters are calculated, denoted as $c_0, c_1,..c_{k-1}$. The next step involves calculating the distances between the test data points $\overline{y_i}$ and the centroids $D_{\overline{y_i},c_0}, D_{\overline{y_i},c_1},..., D_{\overline{y_i},c_{k-1}}$ using the proposed quantum algorithms. The goal is to compare these distances to assign each test data point to the closest cluster, determining $t$ such that $\overline{y_i} \in c_t$. After classification, evaluation metrics are calculated to assess the efficiency and accuracy of the algorithms.
The problem can be formally defined as finding the distances using measurements from quantum circuits designed according to the proposed algorithms:
\begin{eqnarray}
D_{\overline{y_i},c_l}:=M(QC), 
\end{eqnarray}
where \(M(QC)\) represents the measurement outcomes from the quantum circuits.

\begin{figure}[]
\begin{subfigure}{\linewidth}
\includegraphics[width=\linewidth]{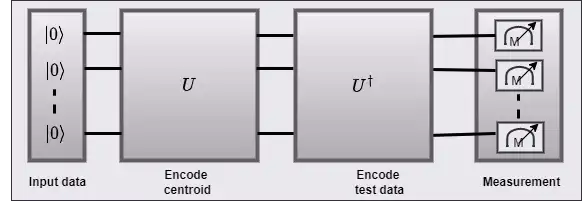}
\caption{}
\label{Fig2a}
\end{subfigure}\hfill
\begin{subfigure}{\linewidth}
\includegraphics[width=\linewidth]{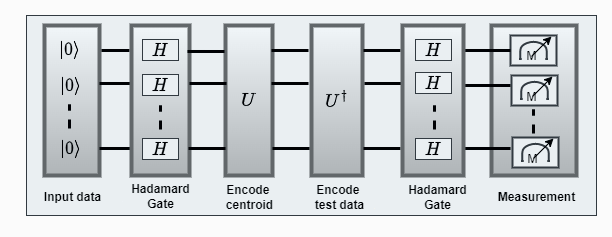}
\caption{}
\label{Fig2b}
\end{subfigure}\hfill
\begin{subfigure}{\linewidth}
\includegraphics[width=\linewidth]{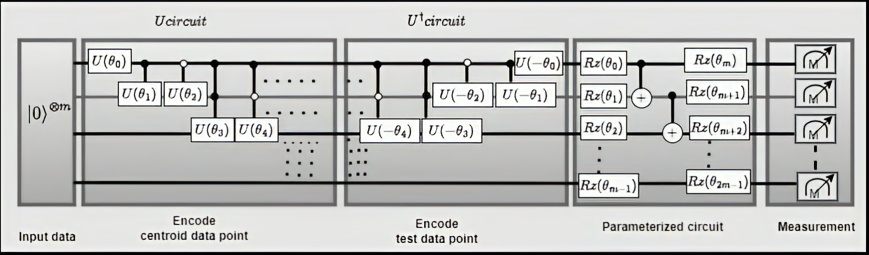}
\caption{}
\label{Fig2c}
\end{subfigure}\hfill
\caption{Quantum Circuit for (a) $UU^{\dagger}$ Algorithm, (b) Variational $UU^{\dagger}$ Algorithm, (c) $UU^\dagger$-QNN Algorithm. }
\label{Fig2}
\end{figure}
\subsection{Quantum Algorithms}

\subsubsection{$UU^\dagger$ Algorithm}
The $UU^\dagger$ method determines the inner product between the centroid and the test data point (see Fig. \ref{Fig2a} and Algorithm \ref{algo1}). The unitary operator $U_1$ encodes the centroid data in the state:
\begin{eqnarray}
\ket{A} = U_1\ket{0}^{\otimes m}
\label{1}
\end{eqnarray}
In this case, the maximum number of m-qubits encodes the centroid data points. Now, another similar unitary operator $U_2$ is used to encode the test data point:
\begin{eqnarray}
\ket{B} = U_2\ket{0}^{\otimes m}
\label{2}
\end{eqnarray}

The inner product between the test data and the centroid is given by:
\begin{eqnarray}
\braket{A|B} = \braket{B|A}&=& ^{m \otimes}\bra{0}U_2^{\dagger}U_1\ket{0}^{\otimes m} \nonumber\\
&=&^{m \otimes}\bra{0}T\ket{0}^{\otimes m} \label{eq_3}
\end{eqnarray}
Here, the operation $T = U_2^{\dagger}U_1$ is performed, which involves applying $U_1$ followed by $U_2^{\dagger}$ to the state $\ket{0}^{\otimes m}$. The operator T is an arbitrary $m$-qubit operator that produces an arbitrary state in the $m$-qubit system when applied to $\ket{0}^{\otimes m}$. In general, the state looks like this:
\begin{eqnarray}
T\ket{0}^{\otimes m} &=& b_0\ket{000...000}+b_1\ket{000...001}\nonumber \\
&+&b_2\ket{000...010}+...+ b_{2^m-1}\ket{111...111}
\label{4}
\end{eqnarray}
Substituting into Eq. \eqref{eq_3}, it becomes:
\begin{eqnarray}
^{m \otimes}\bra{0}T\ket{0}^{\otimes m} &=& b_0
\label{5}
\end{eqnarray}
Here, $b_0$ is the coefficient of state $\ket{000...000}$, which is a real number since it does not have associated phase information (any phase can be taken as common). Furthermore, $b_0$ represents the square root of the probability of measuring the state $\ket{000...000}$:
\begin{eqnarray}
b_0 &=& \sqrt{P_{000...000}}
\label{6}
\end{eqnarray}
Therefore, from Eqs. \eqref{eq_3} and \eqref{6}, the inner product is:
\begin{eqnarray}
\braket{A|B} &=& \sqrt{P_{000...000}} =\sqrt{P_{\ket{0}^{\otimes m}}}
\label{7}
\end{eqnarray}
Thus, the inner product can be calculated by measuring the circuit and determining the square root of the probability of the $\ket{0}^{\otimes m}$ state. The quantum circuit model is initialized with $|0\rangle^{\otimes {m}}$ for data processing, leveraging the controlled $U(\theta_i)$ and anti-controlled unitary $U^\dagger(-\theta_i)$ gates to encode the centroid and test data, respectively. 

\begin{algorithm}
\caption{$UU^\dagger$ algorithm}
\label{algo1}
\begin{algorithmic}[1]
\REQUIRE: Data features and centroid features\;
\ENSURE: Classification of data-points in two clusters\;
\STATE Set all qubits to the $\ket{0}^{\otimes m}$ state

\STATE Encode centroid features in the $U$ operator to produce $\ket{\psi_{0}}$ and $\ket{\psi_{1}}$ representing clusters $C_0$ and $C_1$ respectively
\FOR{$i \in \mathbb{P}$}
    \STATE Encode test data features in $U^\dagger$ to obtain state $\ket{\phi_{i}}$
    
    \STATE Measure the quantum circuit using the first centroid to get ${\langle \phi_{i} | \psi_{0} \rangle}$ and then with the second centroid to get ${\langle \phi_{i} | \psi_{1} \rangle}$ as given in Eq. \ref{eq_3}
    
    \IF{${\langle \phi_{i} | \psi_{1} \rangle > \langle \phi_{i} | \psi_{0} \rangle}$}
        \STATE Corresponding test data belongs to $C_1$
    \ELSE
        \STATE Corresponding test data belongs to $C_0$
    \ENDIF
\ENDFOR
\end{algorithmic}
\end{algorithm}
\begin{algorithm}
\caption{Variational $UU^\dagger$ 
}
\label{algo2}
\begin{algorithmic}[1]

\REQUIRE Data features and centroid features
\ENSURE Classification of data-points in two clusters

\STATE\textbf{Initialization:} Set all qubits to the $\ket{0}^{\otimes m}$ state
\STATE Construct a circuit layer comprising of $n$ Hadamard gates, $N$ $U$ gates and then repeat it $n$ times
\STATE Construct the dagger of the circuit layer and repeat it $n$ times
\STATE Encode the centroid features in the $U$ gates of each layer, resulting in $\ket{\psi_{0}}$ and $\ket{\psi_{1}}$ representing clusters $C_0$ and $C_1$ respectively
\FOR{$i \in \mathbb{P}$}
     \STATE Encode the test data features in the $U^{\dagger}$ gates of each layer, resulting in the state $\ket{\phi_{i}}$
    \STATE Measure the quantum circuit using the first centroid data, followed by a measurement with the second centroid data.
    \IF{${\langle \phi_{i} | \psi_{0} \rangle > \langle \phi_{i} | \psi_{1} \rangle}$}
        \STATE Corresponding test data belongs to $C_0$
    \ELSE
        \STATE Corresponding test data belongs to $C_1$
    \ENDIF
\ENDFOR
\end{algorithmic}
\end{algorithm}
\subsubsection{Variational $UU^\dagger$ Algorithm}
The variational $UU^\dagger$ algorithm extends the standard $UU^\dagger$ technique by incorporating additional layers of Hadamard gates. The primary modification is the placement of a Hadamard layer immediately after initializing the state $\ket{0}^{\otimes m}$ and before the measurement stage, as shown in Fig. \ref{Fig2b} and Algorithm \ref{algo2}. In this algorithm, the operations $(HU)$ and $(U^\dagger$$H)$ are arranged in layers with equal repetitions. Specifically, applying the operation $(HU)$ for $n$ times, followed by $(U^\dagger$$H)$ $n$ times, constructs a variational circuit of layers $n$. 
The variational $UU^\dagger$ algorithm has significant advantages over the $UU^\dagger$ algorithm due to the additional flexibility of the variational layers. These extra layers improve expressibility and enable a better feature transformation, allowing the model to capture more complex patterns in certain datasets. 
After executing the circuit and performing the measurements, the probability of the state $\ket{0}^{\otimes m}$ is determined. The classification decision is based on comparing these probabilities calculated using different centroids. A data point is assigned to the first cluster if the probability $\ket{0}^{\otimes m}$ calculated using the first centroid is higher than that calculated using the second centroid; otherwise, it is assigned to the second cluster. Mathematically, this can be expressed as: if ${\langle \phi_{i} | \psi_{1} \rangle > \langle \phi_{i} | \psi_{0} \rangle}$, then the data point is assigned to cluster 1; otherwise, it belongs to cluster 0.

\subsection{$UU^\dagger$-Quantum Neural Networks}
QNNs utilize variational quantum circuits (VQCs) with multilayer architectures, where entanglement layers replace traditional activation functions. Our work proposes a modified QNN that combines the $UU^{\dagger}$ method with a VQC using tunable parameters. The $UU^\dagger$-QNN consists primarily of a $U$ circuit, a $U^\dagger$ circuit, and a VQC, as shown in Fig. \ref{Fig2c}.
The circuit applies a series of controlled unitary operations $U(\theta_0), U(\theta_1), \dots, U(\theta_m)$, which encode centroid data points into the quantum state, followed by inverse unitary operations $U^\dagger(-\theta_0), U^\dagger(-\theta_1), ….., U^\dagger(-\theta_m)$ to encode test data points, allowing comparison of quantum states. To further process the encoded data, a sequence of parameterized rotations is introduced around the Z axis, denoted $R_z(\theta_0), R_z(\theta_1), \dots,..., R_z(\theta_{2m-1})$. These rotations, parameterized by angles $\theta_i$, are optimized during training to enhance the discriminative power of the model. The variational layers in $UU^\dagger$-QNN were configured with alternating single-qubit rotations ($R_x, R_z$) and entanglement gates (e.g., CNOTs) to ensure efficient exploration of the Hilbert space. Finally, qubits are measured on a computational basis, collapsing the quantum state into a classical bit string that serves as input for downstream analysis or decision-making.
The $U$ circuit includes gates $U$, control $U$, and anti-control $U$, as shown in Fig. \ref{Fig2c}; the number of gates is determined by the number of features in the dataset $N$. The features of the centroids are encoded in the parameters $\theta$ of the gates $U$, with $\phi$ and $\lambda$ set to zero. Similarly, the $U^\dagger$ circuit consists of $U^{\dagger}$, control $U^{\dagger}$, and anti-control $U^{\dagger}$ gates, where the features of the test data are encoded in the $\theta$ parameters, with $\phi$ and $\lambda$ set to zero. 
Then, VQCs, hybrid quantum-classical approaches that take advantage of both types of computation, are applied to the quantum system. 
Unlike conventional parameterized quantum circuits that suffer from barren plateaus and gradient decay during the training process, the $UU^\dagger$-QNN cost function inherently preserves unitary consistency and ensures stable coherence and noise resilience, which is reliable for a noisy high-dimensional quantum environment. 
These circuits feature controllable parameters that are iteratively optimized by a classical computer.
The loss function used in the $UU^\dagger$-QNN algorithm is defined as:
\begin{eqnarray}
\text{Loss}= (1-\text{Accuracy})^2
\label{eq_22}
\end{eqnarray}
\begin{algorithm}
\caption{Optimization Algorithm}
\label{algo6}
\begin{algorithmic}[1]
\REQUIRE Initial parameters $params$, Number of iterations $num\_iterations$
\ENSURE Plot of accuracy vs iteration
\STATE Define the Accuracy function:
\STATE Define the Loss function:
\STATE Define the optimization procedure:
\STATE \textbf{Function} {Optimize\_Params} ($initial\_params$, $num\_iterations$)
    \STATE Initialize empty lists: $accuracy\_list$, $iteration\_list$
    \STATE Initialize $optimize\_IP$ as None
    \FOR{$i\leftarrow 1$ to $num\_iterations$}
        \IF{$i = 1$}
            \STATE $optimize\_IP \leftarrow$ COBYLA optimization on Loss() using $initial\_params$
        \ELSE
            \STATE $optimize\_IP \leftarrow$ COBYLA optimization on Loss() using $optimize\_IP.x$
        \ENDIF
        \STATE Calculate accuracy using $\text{Accuracy\_function}(optimize\_IP.x)$
        \STATE Append accuracy to $accuracy\_list$
        \STATE Append $i$ to $iteration\_list$
    \ENDFOR
\STATE Plot $accuracy\_list$ against $iteration\_list$\;
\end{algorithmic}
\end{algorithm}
The loss function is optimized using the COBYLA optimizer to find the optimal parameters for the quantum circuit (Algorithm \ref{algo6}).
The above loss function was chosen to emphasize maximizing accuracy while strongly penalizing misclassifications, offering a smooth gradient that aids convergence with optimizers such as COBYLA. Its simplicity and computational efficiency make it well-suited for quantum circuits.
The input to the loss function consists of the parameterized quantum circuit, which uses standard quantum gates such as single-qubit rotation gates $Rx(\alpha_j)$, $Ry(\beta_j)$ and $Rz(\gamma_j)$, controlled-Not (CNOT) gates, and controlled-Z (CZ) gates. In our implementation, $Rz(\theta)$ and the CNOT gates are used to entangle within the circuit. The detailed $UU^\dagger$-QNN method is provided in Algorithm \ref{algo3} and the definitions of the rotation gates $R_{x}$$(\alpha)$, $R_{y}$$(\beta)$, $R_{z}$$(\gamma)$, CNOT (CX) and CZ are presented as follows.
\begin{eqnarray}
R_{x}(\alpha) &=&\cos{\alpha}\ket{0}\bra{0}- i\sin{\alpha}\ket{0}\bra{1}\nonumber\\
&&-i\sin{\alpha}\ket{1}\bra{0}+\cos{\alpha}\ket{1}\bra{1},\nonumber\\
R_{y}(\beta) &=&\cos{\beta}\ket{0}\bra{0} - \sin{\beta}\ket{0}\bra{1}\nonumber\\
&&+\sin{\beta}\ket{1}\bra{0}+ \cos{\beta}\ket{1}\bra{1},\nonumber\\
R_{z}(\gamma) &=&e^{-i\gamma}\ket{0}\bra{0}+e^{i\gamma}\ket{1}\bra{1},\nonumber\\
CX &=& \ket{00}\bra{00}+\ket{01}\bra{01}+\ket{10}\bra{11}+\ket{11}\bra{10},\nonumber\\
CZ &=& \ket{00}\bra{00}+\ket{01}\bra{01}+\ket{10}\bra{10}-\ket{11}\bra{11}.\nonumber
\label{8}
\end{eqnarray}
Regarding the variational layers, our quantum circuits are designed with parameterized single-qubit rotations followed by entangling gates (e.g., CNOT), repeated across layers to form a variational ansatz. These layers are optimized via gradient-based classical routines to minimize the loss function. The parameters of these gates allow the model to learn complex representations, analogous to weights in classical neural networks. Including variational layers improves the quantum model's capacity to approximate non-linear transformations, which is critical for learning complex data distributions in generative tasks.
\begin{algorithm}
\caption{$UU^\dagger$-QNN for Classification}
\label{algo3}
\begin{algorithmic}[1]
\REQUIRE Features of data points, Features of centroids, initial parameter list $params$
\ENSURE Plot of accuracy against iteration

\STATE Initialize all $m$ qubits in the quantum circuit in $\ket{0}^{\otimes m}$ state.

\STATE Encode the centroid data in the $U$ circuit, resulting in quantum states $\ket{\text{Cluster}_1}$ and $\ket{\text{Cluster}_0}$
\STATE Encode test datapoint in the $U^{\dagger}$ circuit, resulting in the state $\ket{\text{Test}_i}$.

\STATE \textbf{Variational Quantum Circuit}:
\STATE \textbf{Layer 1}:
\FOR{$j = 0$ to $m-1$}
\STATE Apply $R_z(\theta = params[j])$ gate to qubit $j$
\ENDFOR
\STATE Apply one hidden layer of $m$ CNOT gates between each pair of nearest neighbor qubits.

\STATE \textbf{Layer 2}:
\FOR{$j = 0$ to $m-1$}
\STATE Apply $R_z(\theta = params[j+m])$ gate to qubit $j$
\ENDFOR

\STATE Perform optimization using Algorithm \ref{algo6} and get the accuracy plots.
\end{algorithmic}
\end{algorithm}
\section{Experimental Results\label{SecIII}}
\subsection{Datasets}
The proposed quantum algorithms are evaluated using four datasets that provide various data types to assess 5G-enabled IoMT systems. The 5G-SA dataset contains 2,835 rows and 14 columns generated by the NYUSIM 3.0 millimeter wave simulator, which incorporates atmospheric parameters relevant to South Asia \cite{q_26}. 
It has key features such as transmitter-receiver distance, frequency band, environmental obstacles, and received signal strength indicator (RSSI), with missing RSSI values addressed through regression-based imputation and noisy data smoothed using Kalman filters.
The L5G1.0 dataset includes 68,118 rows and 19 columns of 5G throughput data from a 1,300-meter loop near the U.S. Bank Stadium in Minneapolis, covering urban infrastructure \cite{q_23, q_27}. 
It includes latency, throughput, packet loss, jitter, and signal-to-noise ratio (SNR) as features, where missing data were handled using forward/backward filling techniques, and outliers were mitigated using Z-score and interquartile range (IQR) methods to ensure robust network performance analysis.
The WE20 dataset features 44 attributes, with 35 network flow metrics and eight biometric features, collected from a real-time healthcare monitoring testbed, making it ideal for evaluating quantum algorithms in healthcare-focused IoMT networks \cite{WE20_works, q_28}.
It has biometric signals such as heart rate, ECG, and accelerometer readings, along with contextual metadata as features. Missing data was addressed through interpolation and median imputation, while noisy signals were cleaned using wavelet transforms and low-pass filters to preserve physiological patterns.
The PS-IoT dataset includes 97606 rows and 5 column features such as device identity, timestamp of activity, data sensitivity, privacy, and threat level of IoT security incidents \cite{q_229}.
\subsection{Preprocessing}
All unnecessary and irrelevant columns were removed during pre-processing. Then, 14 columns from the 5G-SA dataset, 13 columns from the L5G1.0 dataset, 33 columns from the WE20 dataset, and 5 columns from the PS-IoT dataset were considered. The feature values of all datasets were normalized to the range $[0, \pi]$, and data points were separated into two clusters per dataset using the $k$-means clustering method. The quantum circuit initializes with the state $\ket{0}^{\otimes m}$, where the number of qubits $m$ is selected based on the dataset's dimensionality and complexity. For example, the 5G-SA dataset, which involves high-dimensional path loss metrics, used $m = 6$ qubits; the L5G1.0 dataset, which encodes network-level metrics, required $m = 5$ qubits; the WE20 dataset, which includes biometric signals, used $m = 7$ qubits; and $m = 6 $ qubits for PS-IoT is needed, which includes timestamp, hour, minute, day, month and weekend. The qubit count was determined by balancing comprehensive feature encoding with the practical limits of quantum hardware.

\begin{figure*}[]
\centering
\begin{subfigure}{0.5\linewidth}
\centering
\includegraphics[width=0.8\linewidth]{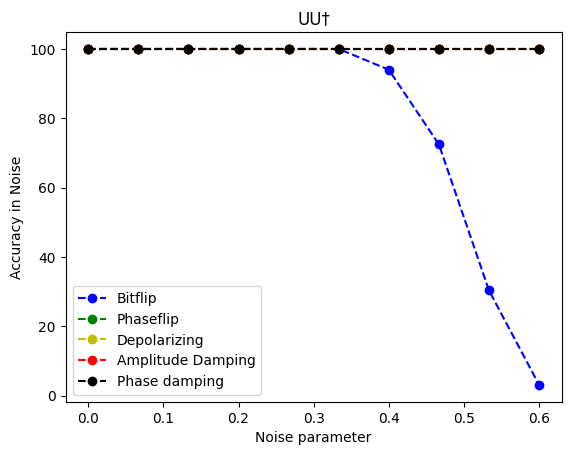}
\caption{}
\label{Fig3a}
\end{subfigure}\hfill
\begin{subfigure}{0.5\linewidth}
\centering
\includegraphics[width=0.8\linewidth]{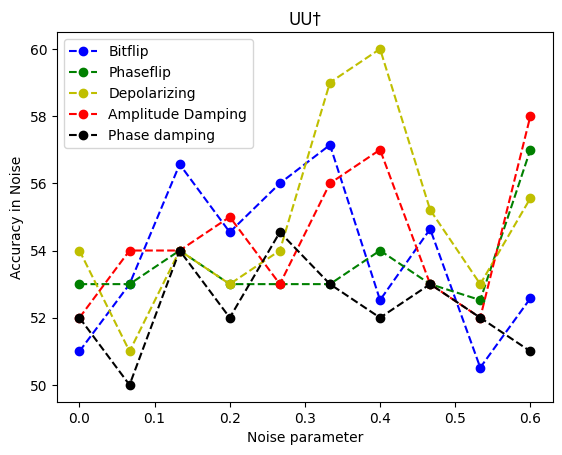}
\caption{}
\label{Fig3e}
\end{subfigure}\hfill
\begin{subfigure}{0.5\linewidth}
\centering
\includegraphics[width=0.8\linewidth]{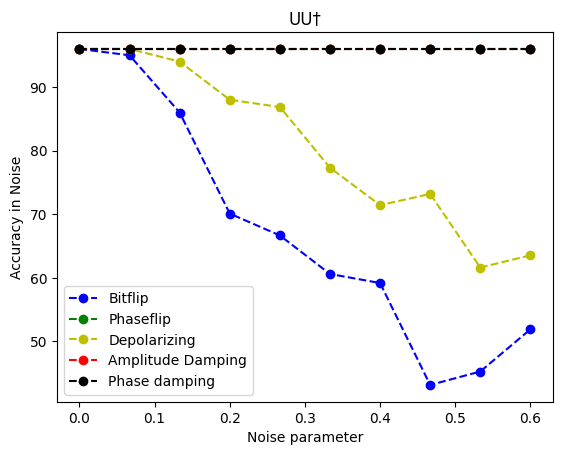}
\caption{}
\label{Fig3i}
\end{subfigure}\hfill
\begin{subfigure}{0.5\linewidth}
\centering
\includegraphics[width=0.7\linewidth]{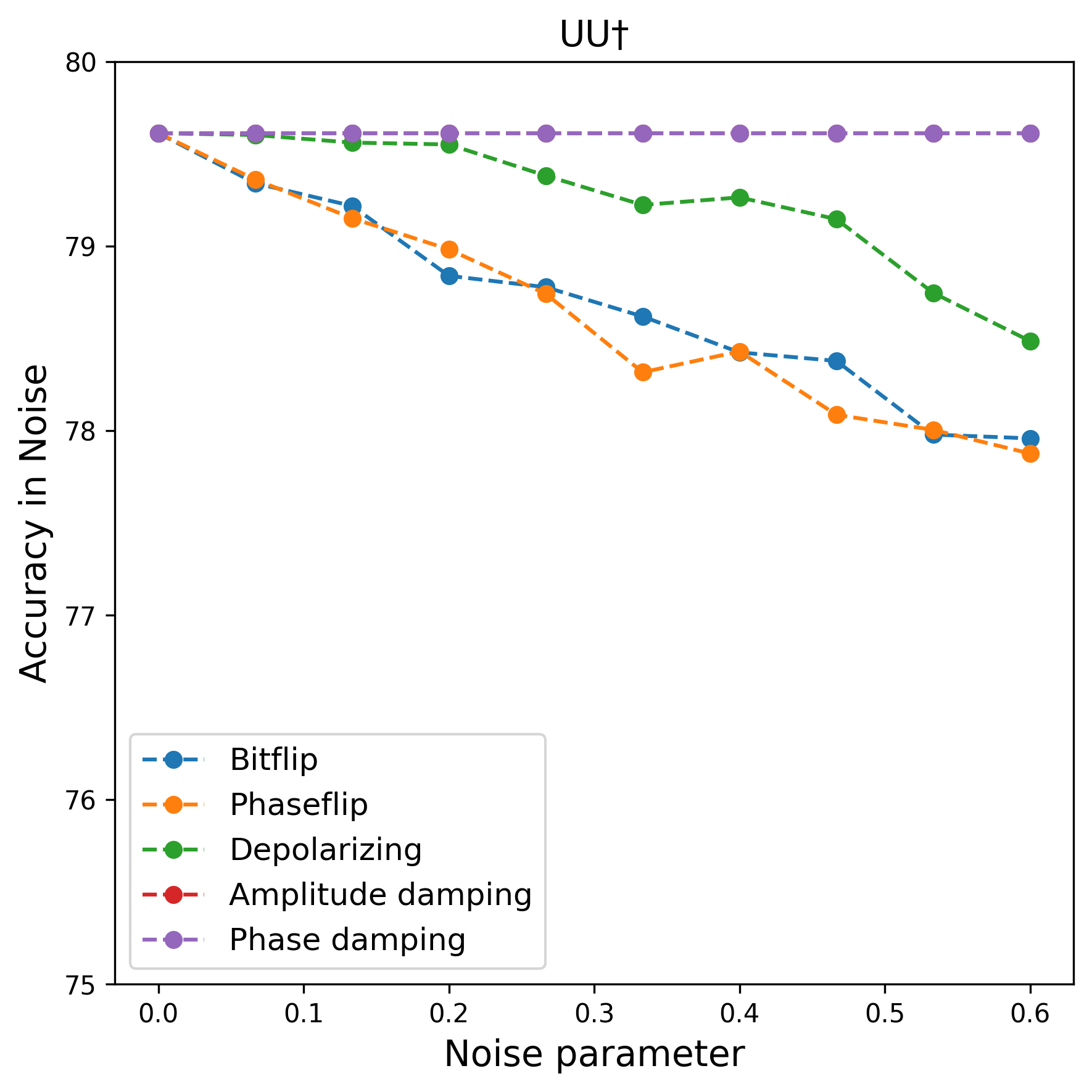}
\caption{}
\label{Fig3i}
\end{subfigure}\hfill
\caption{The Accuracy of $UU^\dagger$ Using the $k$-means Clustering in a Noisy Environment for (a) 5G-SA, (b) L5G1.0, (c) WE20 and (d) PS-IoT}.
\label{Fig3}
\end{figure*}

\subsection{Metrics, Hyperparameters and Noise Models}
In multiclass classification, key evaluation metrics include accuracy, precision, recall, and the F1 score, providing information on general and class-specific model performance. The proposed quantum algorithms were implemented using the IBM QASM simulator with 1000 shots. The COBYLA optimizer was used for the $ UU^ {\dagger}$-QNN algorithm, with all initial parameters set to 0.1 for the datasets.
Robustness and efficiency were evaluated against five key quantum noise models, which are essential to indicate error correction and improve computational reliability \cite{sritam23}. The selected noise models are bit-flip, phase-flip, depolarizing, amplitude damping, and phase damping, which simulate realistic interference and degradation. Bit-flip and phase-flip capture discrete qubit state errors, often caused by electromagnetic interference (EMI) from medical devices or unstable wireless environments. Depolarizing noise represents random decoherence due to hardware imperfections, amplitude and phase damping models capture energy loss and dephasing effects resulting from environmental interactions, such as photon dissipation or oscillator instability.


\begin{figure*}[]
\begin{subfigure}{.5\textwidth}
\centering
\includegraphics[width=0.8\linewidth]{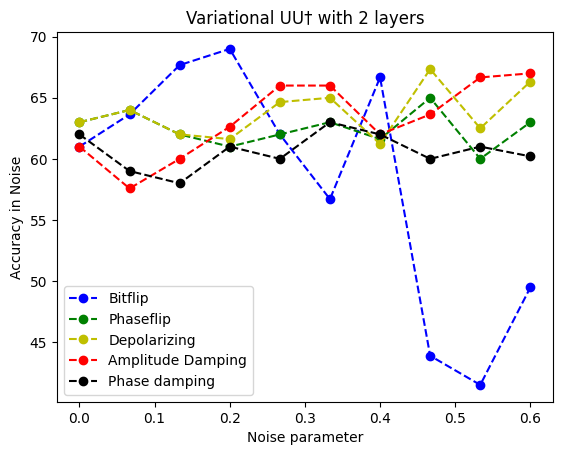}
\caption{}
\label{Fig3b}
\end{subfigure}\hfill
\begin{subfigure}{.5\textwidth}
\centering
\includegraphics[width=0.8\linewidth]{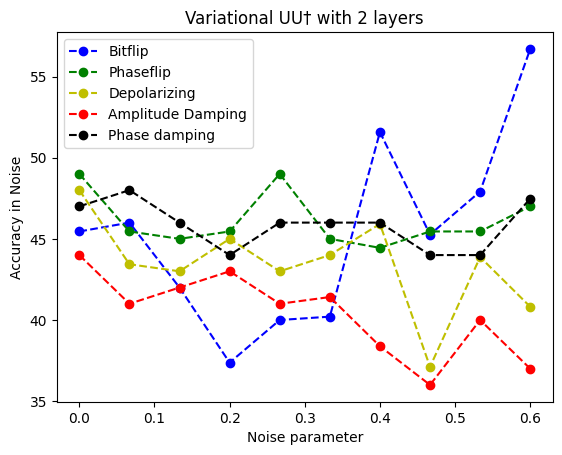}
\caption{}
\label{Fig3f}
\end{subfigure}\hfill
\begin{subfigure}{.5\textwidth}
\centering
\includegraphics[width=0.8\linewidth]{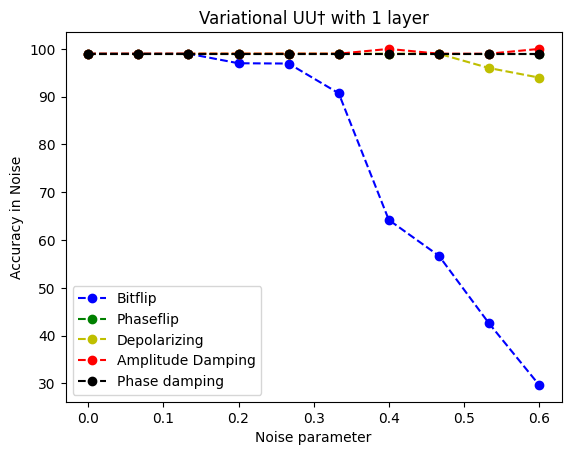}
\caption{}
\label{Fig3j}
\end{subfigure}\hfill
\begin{subfigure}{.5\textwidth}
\centering
\includegraphics[width=0.7\linewidth]{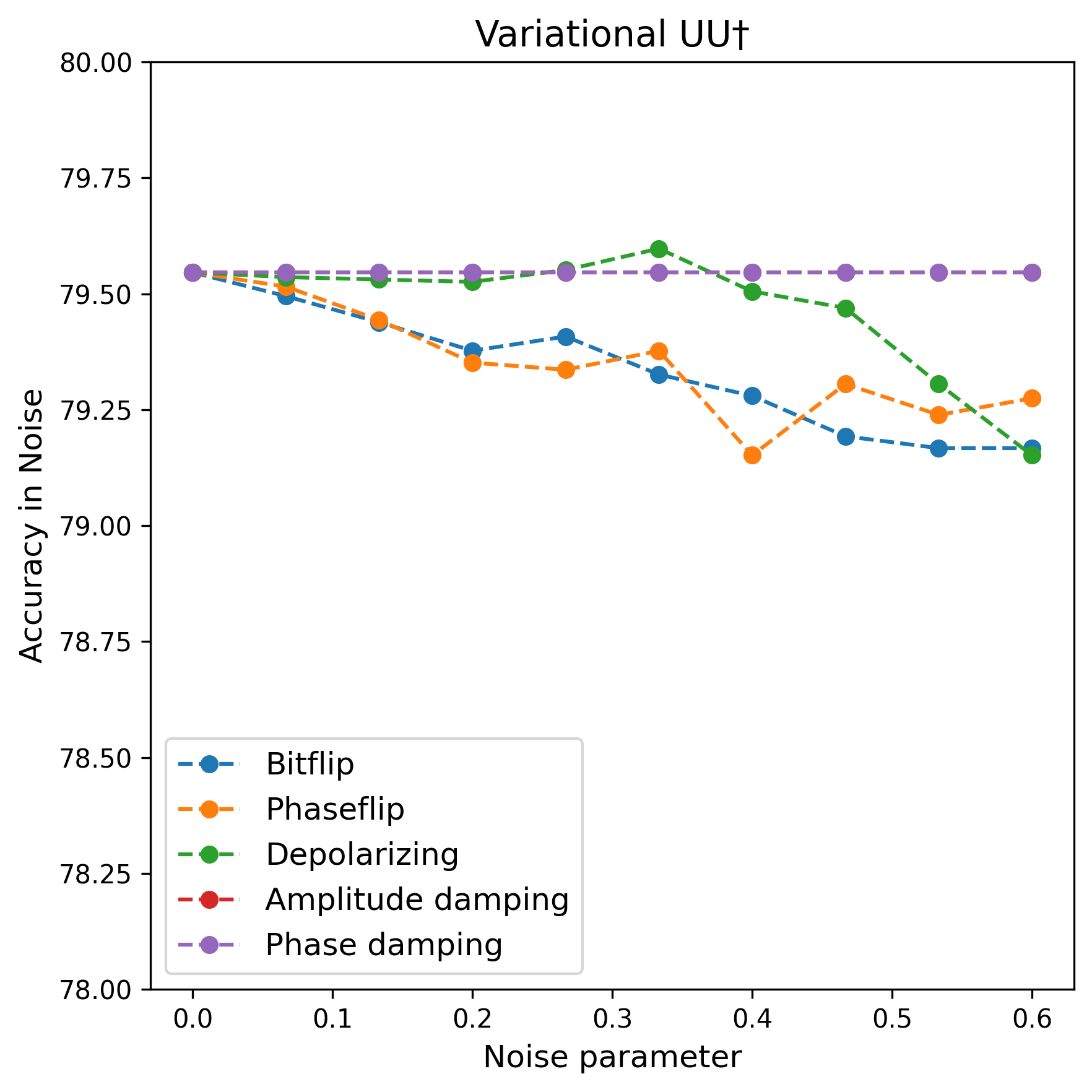}
\caption{}
\label{Fig3j}
\end{subfigure}\hfill
\caption{
Accuracy Performance of the Variational  $UU^\dagger$ Method under a Noisy Environment for the (a) 5G-SA, (b) L5G1.0, (c) WE20, (d) PS-IoT datasets.
}
\label{Fig3}
\end{figure*}

\begin{figure*}[]
\begin{subfigure}{0.5\textwidth}
\centering
\includegraphics[width=0.8\linewidth]{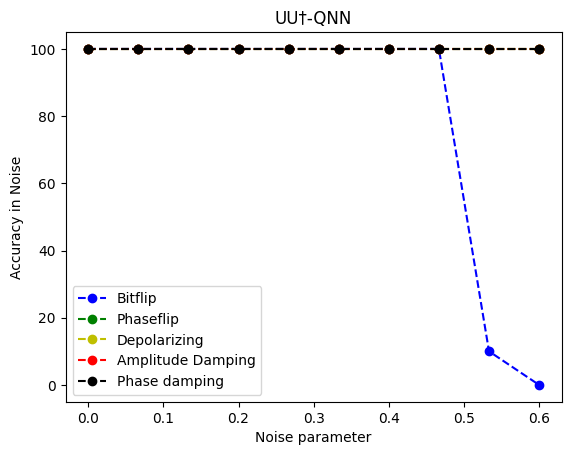}
\caption{}
\label{Fig3c}
\end{subfigure}\hfill
\begin{subfigure}{0.5\textwidth}
\centering
\includegraphics[width=0.8\linewidth]{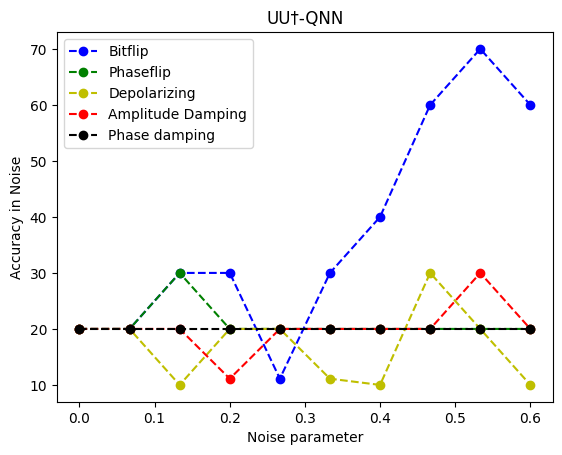}
\caption{}
\label{Fig3g}
\end{subfigure}\hfill
\begin{subfigure}{.5\textwidth}
\centering
\includegraphics[width=0.8\linewidth]{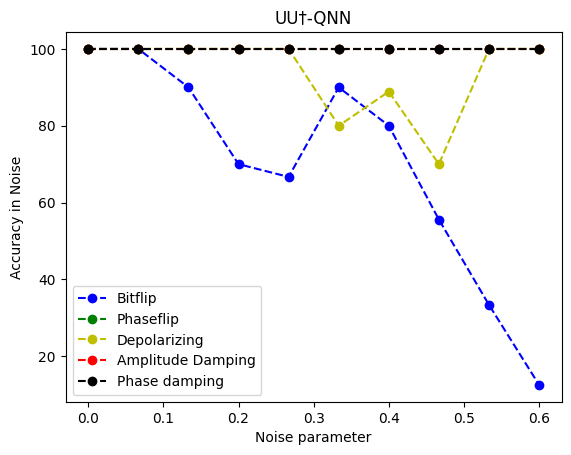}
\caption{}
\label{Fig3k}
\end{subfigure}\hfill
\begin{subfigure}{.5\textwidth}
\centering
\includegraphics[width=0.7\linewidth]{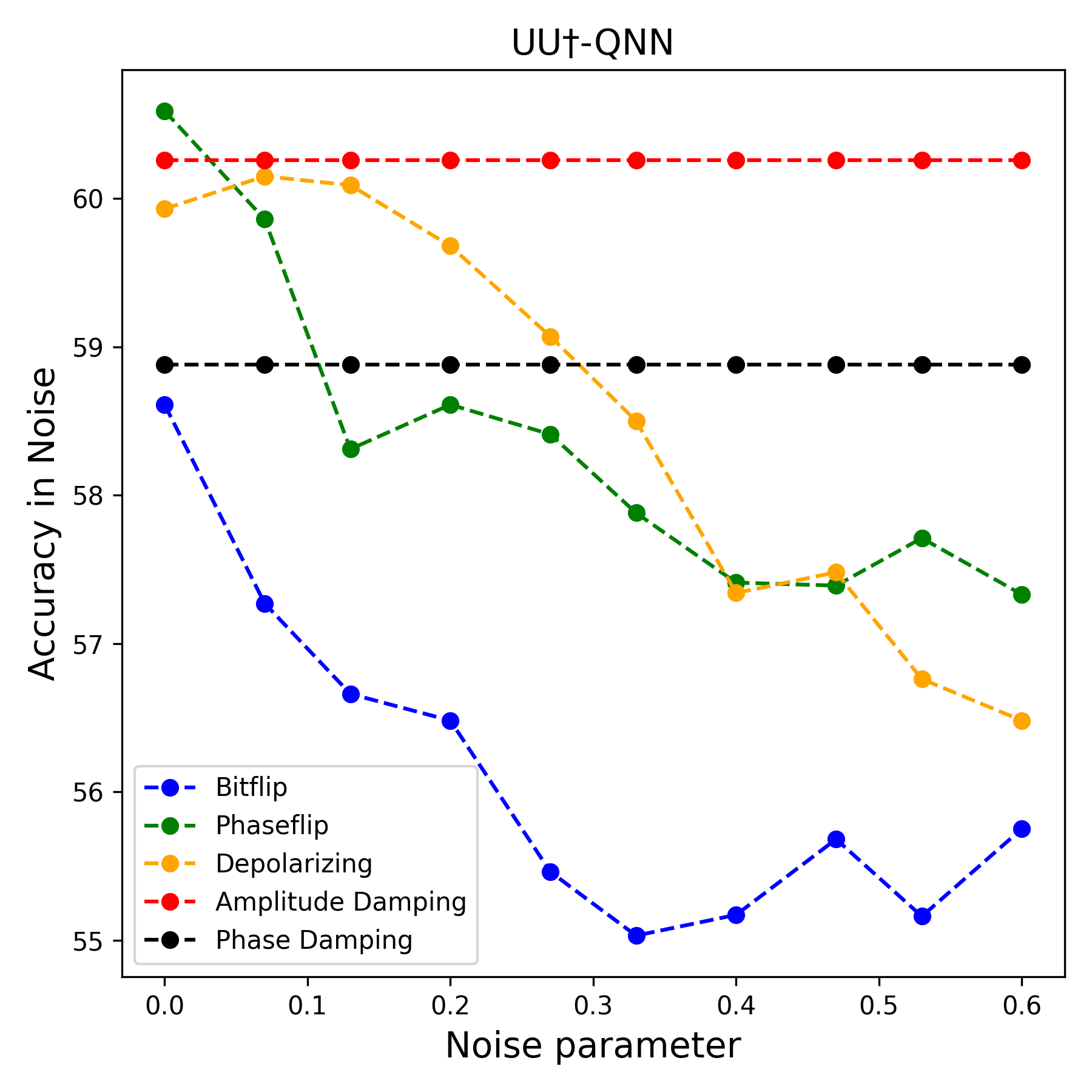}
\caption{}
\label{Fig3k}
\end{subfigure}\hfill
\caption{
Accuracy Performance of the $UU^\dagger$-QNN Method under a Noisy Environment for the (a) 5G-SA, (b) L5G1.0, (c) WE20, and  (d) PS-IoT datasets.}
\label{Fig3}
\end{figure*}
\subsection{Results of Quantum Algorithms}
\subsubsection{$UU^\dagger$ Algorithm}
In the $UU^\dagger$ algorithm, the $k$-means clustering method is used first to obtain the centroids. Then a quantum circuit is constructed, where the characteristic values of the centroids are encoded in the $\theta$ parameters of the $U$ gates, while $\phi$ and $\lambda$ are set to zero. The test data features are similarly encoded into $U^\dagger$ gates but in reverse order (Fig.~\ref{Fig2c}). For each data point, the quantum circuit is executed and measured, and the probability of measuring the state $\ket{0}^{\otimes m}$ is used to compute the inner product between the test data and each centroid (Eq.~\ref{7}). This process is repeated for both centroids, and each test data point is assigned to the cluster corresponding to the centroid that produces the highest probability, as outlined in Algorithm~\ref{algo1}. This procedure is applied to all four datasets, and each data point is classified and labeled accordingly. The achieved accuracies are 73.09\% for the 5G-SA dataset, 52.26\% for the L5G1.0 dataset, 96\% for the WE20 dataset, and 54\% for the PS-IoT dataset.

\begin{table}
  \centering
  \renewcommand{\arraystretch}{1.2} 
  \begin{tabular}{|c|c|c|c|c|c|}
    \hline
    \textbf{Dataset} & \textbf{Algorithm} & \textbf{Accuracy} & \textbf{Precision} & \textbf{Recall} & \textbf{F1 Score} \\
    \hline
    \multirow{5}{*}{5G-SA} & SVM & 0.85 & 0.85 & 0.85 & 0.85 \\
    & ANN & 0.86 & 0.87 & 0.87  & 0.86 \\
    & $UU^\dagger$ & 0.73  & 1.00 & 0.65  & 0.78\\
    & V-$UU^\dagger$ &0.72 & 0.46 & 1.00 & 0.63 \\
    & $UU^\dagger$-QNN & 1.00 &0.00 & --& -- \\
    \hline
    \multirow{5}{*}{L5G1.0} & SVM & 0.65 & 0.65 & 0.65  & 0.64 \\
    & ANN & 0.78 & 0.78 & 0.78 & 0.78\\
    & $UU^\dagger$ & 0.52 & 1.00 & 0.52  & 0.69\\
    & V-$UU^\dagger$& 0.54  & 1.00 & 0.55 & 0.71\\
    & $UU^\dagger$-QNN & 0.20 & 0.00 &-- & -- \\
    \hline
    \multirow{5}{*}{WE20} & SVM & 0.99 & 0.99 & 0.99  & 0.99 \\
    & ANN & 1.00 & 0.87 & 0.87 & 0.86\\
    & $UU^\dagger$ & 0.96 & 1.00 & 0.33  & 0.50\\
    & V-$UU^\dagger$ & 0.99 & 1.00 & 0.83  & 0.91\\
    & $UU^\dagger$-QNN & 1.00 & 1.00 & 1.00  & 1.00\\
    \hline
        \multirow{5}{*}{PS-IOT} & $ SVM $ & 0.81 & 0.82 & 0.97 & 0.89\\
     & $ANN$ & 0.82 & 0.84 & 0.96 & 0.89\\
    & $UU^\dagger$ & 0.54 & 0.82 & 0.54  & 0.65\\
    & V-$UU^\dagger$ & 0.80 & 0.80 & 0.99  & 0.89\\
    & $UU^\dagger$-QNN & 0.55 & 0.80 & 0.58  & 0.67\\ 
    \hline
  \end{tabular}
  \caption{Comparison of Performance Metrics for 5G-SA, L5G1.0, WE20, and PS-IoT. Variational $UU^\dagger$ (V-$UU^\dagger$).}
  \label{table5}
\end{table}
\subsubsection{Variational $UU^\dagger$ Algorithm}
In the variational $UU^\dagger$ algorithm, centroids are obtained using $k$-means clustering, followed by applying Hadamard, unitary $U$, and $U^\dagger$ circuits to encode the data. The probability of the $\ket{0}^{\otimes m}$ state is measured to compute the inner product between the centroids and the test data points, which determines the cluster assignments. The process is repeated in multiple layers, from one to ten. 
The performance of the variational $UU^\dagger$ algorithm is improved due to additional variational layers that capture complex data patterns for specific datasets, with extra feature handling. The accuracies achieved are 72\% for the 5G-SA dataset, 54\% for the L5G1.0 dataset, 99\% for the WE20 dataset, and 80\% for the PS-IoT dataset.

\subsubsection{$UU^\dagger$-Quantum Neural Networks}
The $UU^\dagger$-QNN algorithm iteratively calculates the loss function for 10 data points from each dataset, optimizing the parameters in each step using the COBYLA optimizer. Accuracy is evaluated by comparing the output of the variational quantum circuit with the classical dot-product results. The VQC outputs are simulated using the IBM QASM simulator with 1000 shots, and the optimization runs for 20 iterations, tracking accuracy at each epoch. The results show that for the 5G-SA dataset, the accuracy remains consistent at 100\%; for L5G1.0, it stays at 20\%; for the WE20 dataset, accuracy also remains at 100\% and maintains the accuracy at 57\% for the PS-IoT dataset. A comparison of the accuracies of $UU^\dagger$, variational $UU^\dagger$, and $UU^\dagger$-QNN (Table \ref{table5}) indicates that for the 5G-SA dataset, $UU^\dagger$-QNN outperforms the other methods. For L5G1.0 and PS-IoT datasets, variational $UU^\dagger$ achieves higher accuracy, while for the WE20 dataset, variational $UU^\dagger$ and $UU^\dagger$-QNN surpass $UU^\dagger$.
The low F1 score (0.20) of $UU^{\dagger}$-QNN on the L5G1.0 dataset highlights the need for improved encoding, error mitigation, and more effective optimization to handle complex, noisy, and imbalanced real-world data. 

\subsubsection{Noisy Results}
The robustness of the three quantum algorithms-$UU^\dagger$, variational $UU^\dagger$ and $UU^\dagger$-QNN was evaluated under increasing noise probabilities (\( \gamma = 0 \) to \( 0.6 \)) in the 5G-SA, L5G1.0, 
WE20, and PS-IoT datasets (Fig. 3 - Fig. 5). For the 5G-SA dataset, the $ UU^\dagger$ algorithm degrades to a lower noise level of \( \gamma = 0.35 \) under bit-flip noise while maintaining stable or fluctuating accuracy under other noise models (Fig. 3a). The variational $UU^\dagger$ algorithm (Fig. 4a) shows a similar decreasing trend under bit-flip noise, with a slight recovery at \( \gamma = 0.533 \), and fluctuating accuracy between 55\% and 65\% for other noise types. In contrast, the $UU^\dagger$-QNN algorithm demonstrates superior robustness (Fig. 5a), preserving 100\% accuracy under most noise models and remaining stable up to \( \gamma = 0.48 \) under bit-flip noise.

For the L5G1.0 dataset, both the $UU^\dagger$ and the variational $UU^\dagger$ algorithms exhibit non-robust performance across all noise channels (Figs. 3b and 4b), with significant degradation observed in \( \gamma = 0.4 \) (depolarizing noise) and $ \gamma = 0.08$, and $0.4$ (bit-flip noise). The $UU^\dagger$-QNN algorithm maintains stable accuracy under phase damping. Still, it displays oscillatory and unstable behavior under other noise models, particularly bit-flip noise (Fig. 5b). For the WE20 dataset, the $UU^\dagger$ algorithm’s accuracy decreased to \( \gamma = 0.08 \) under both bit-flip and depolarizing noise (Fig. 3c). The variational $UU^\dagger$ algorithm shows gradual degradation, beginning at \( \gamma = 0.14 \) for bit-flip and worsening up to \( \gamma = 0.48 \) for depolarizing noise (Fig. 4c). In contrast, $UU^\dagger$ -QNN achieves the highest robustness, with a bit-flip accuracy that gradually decreases from \( \gamma = 0.06 \) to \( 0.27 \) and drops to 10\% at \( \gamma = 0.6 \). Under depolarizing noise, the accuracy initially decreases to \( \gamma = 0.28 \), but recovers to 100\% as the noise increases, highlighting the adaptability of the algorithm (Fig. 5c).
For the PS-IoT dataset, the $UU^\dagger$ and variational $UU^\dagger$ algorithms achieve robust accuracy in the phase damping channel while degradation occurs for depolarization in \(\gamma = 0.35\) to \(0.6 \) and the remaining noise models exhibit a gradual decrease in accuracy with \(\gamma = 0\) to \(0.6 \) as shown in Figs. 3d and 4d. The performance of the $UU^\dagger$-QNN algorithm against amplitude damping and phase damping is robust at different accuracies, while the depolarizing noise model decreases in accuracy from \(\gamma = 0.05\) to \(0.6\) and other noise models show a gradual decrease in accuracy, as shown in Fig. 5d.
\subsubsection{Comparative Analysis}
The performance analysis of the proposed quantum algorithms compared to the classical models is shown in Table \ref{tab:my_label_3}. Although quantum models generally outperformed classical methods, the results varied between datasets. For the 5G-SA dataset, the classical method achieved an \( R^2 \) score of 0.89, while the proposed \( UU^\dagger \)-QNN algorithm achieved 1.00 accuracy. In the L5G1.0 dataset, the classical model reported in \cite{q_23} achieved an F1 score of 0.78, while the variational algorithm \( UU^\dagger \) achieved a comparable F1 score of 0.71. In the WE20 dataset, the SVM model from \cite{WE20_works} reached 0.92 accuracy, while the proposed \( UU^\dagger \)-QNN achieved a perfect score of 1.00 on all metrics, outperforming classical baselines.
\textcolor{blue}{}
\begin{table}[]
    \centering
    \begin{tabular}{|c|c|c|}
    \hline
     \textbf{Datasets}    & \textbf{Models} & \textbf{Metrics} \\
     \hline
     5G-SA   & L \cite{q_23} &0.89 (R2)\\
         \hline
    5G-SA   & $UU^{\dagger}$-QNN & 1.00 (Accuracy)\\
         \hline
      L5G1.0   & L \cite{q_23} &0.78 (F1 Score)\\
         \hline
         L5G1.0   & V-$UU^{\dagger}$  &0.71 (F1 Score)\\
         \hline
       WE20  & SVM \cite{WE20_works} & 0.92 (Accuracy)\\
         \hline
         WE20  & $UU^{\dagger}$-QNN & 1.00 (Accuracy)\\
         \hline
    \end{tabular}
    \caption{Comparison of Performance Metrics Between Existing Algorithms and the Proposed Method.}
    \label{tab:my_label_3}
\end{table}

\section{Discussion and Conclusion\label{SecIV}}
Since Industry 5.0 healthcare systems increasingly rely on real-time monitoring and communication with limited power resources, classical machine learning methods face significant limitations in meeting the scalability and efficiency demands of 5G-enabled IoMT environments.  To overcome these challenges, we proposed a classical-quantum hybrid framework incorporating $UU^{\dagger}$, variational $UU^{\dagger}$, and $UU^{\dagger}$-QNN, augmented by $k$-means clustering. This framework demonstrated improvements in energy-efficient processing, with $UU^{\dagger}$-QNN achieving 100\% accuracy in the 5G-SA and WE20 datasets. The quantum algorithms’ robustness was evaluated against five noise models and demonstrated strong performance, although bitflip noise remained the most disruptive. The $UU^{\dagger}$-QNN algorithm remained resilient to most noise models while outperforming classical machine learning in accuracy and energy efficiency, especially for large datasets, and offers a scalable alternative to classical models by addressing exponential memory growth.

In real-world healthcare applications, our developed quantum algorithms have significant potential to transform personalized medicine by enabling early disease prediction, optimized treatment planning, and cost-effective diagnostics, where $UU^{\dagger}$-QNN can detect complex patterns and variational $UU^{\dagger}$ can manage large-scale data to tailor individual patient care. However, decoherence, hardware infidelities, and qubit topology limitations challenge practical deployment on current quantum hardware. Furthermore, variational training suffers from barren plateaus, and the $UU^{\dagger}$-QNN circuits amplify noise on sparse topologies. Our future work will address identified noise vulnerabilities, particularly bit-flip noise, by incorporating advanced error mitigation and quantum error correction methods. Furthermore, we are developing optimized hybrid classical-quantum frameworks and experimentally validating them on near-term quantum devices to enhance the scalability and reliability of Industry 5.0 healthcare applications.

\section{References\label{SecV}}
\bibliographystyle{IEEEtran}

\bibliography{IEEE}

\vfill

\end{document}